\documentclass[aps,pra,twocolumn,groupedaddress]{revtex4-2}

\usepackage[utf8]{inputenc}
\usepackage{float}
\usepackage{graphicx}
\usepackage{amsmath}

\begin{document}
	
	
	\title{Interorbital Interactions in an SU(2)$\otimes$SU(6)-Symmetric Fermi-Fermi Mixture}
	
	

	\author{B.~Abeln*$^1$}
	\author{K.~Sponselee*$^1$}
	\author{M.~Diem$^1$}
	\author{N.~Pintul$^1$}
	\author{K.~Sengstock$^{1,2}$}
	\author{C.~Becker$^{1,2}$}
	
	\affiliation{$^1$Center for Optical Quantum Technologies, University of Hamburg, Luruper Chaussee 149, 22761 Hamburg, Germany}
	\affiliation{$^2$Institute for Laser Physics, University of Hamburg, Luruper Chaussee 149, 22761 Hamburg, Germany}
	\affiliation{*These authors contributed equally to this work}
	
	
	\date{\today}
	
	\begin{abstract}
		
		We characterize inter- and intraisotope interorbital interactions  between atoms in the $ ^{1}\text{S}_{0} $ ground state and the  $ ^{3}\text{P}_{0} $ metastable state in interacting Fermi-Fermi mixtures of $ ^{171} $Yb and $ ^{173} $Yb.
		We perform high-precision clock spectroscopy to measure interaction-induced energy shifts in a deep 3D optical lattice and determine the corresponding scattering lengths.
		We find the elastic interaction of the interisotope mixtures $^{173}$Yb$_\text{e}$-$^{171}$Yb$_\text{g}$ and $^{173}$Yb$_\text{g}$-$^{171}$Yb$_\text{e}$ to be weakly attractive and very similar, while the corresponding two-body loss coefficients differ by more than two orders of magnitude.
		By comparing different spin mixtures we experimentally demonstrate the  SU$(2)\otimes $ SU$(6) $ symmetry of all elastic and inelastic interactions. 
		Furthermore, we measure the spin-exchange interaction in $^{171}$Yb and confirm its previously observed antiferromagnetic nature.
	\end{abstract}
	
	
	\maketitle
	
	\section{Introduction}
	\label{sec:introduction}
	
	Mixtures of ultracold atomic gases offer the opportunity of designing and investigating exotic quantum many-body physics in yet unexplored regimes.
	In the past decades a large variety of different systems involving  mixtures of atoms in different hyperfine states \cite{Myatt1997, Hall1998,Stenger1998,Regal2003}, different isotopes  \cite{Truscott2001,Schreck2001,Papp2008,Fukuhara2009,Tey2010}, different elements \cite{Hadzibabic2002,Stan2004,Inouye2004, Ospelkaus2006a,Taglieber2008,Wille2008,Hara2011} and different orbital states \cite{Scazza2014a,Cappellini2014b,Zhang2014} have been studied, aiming at the creation of e.g.weakly bound heteronuclear molecules \cite{Ospelkaus2006b}, dipolar molecules \cite{Ospelkaus2008,DeMiranda2011,Chotia2012} and polaronic quasiparticles \cite{Kohstall2012,Koschorreck2012,Massignan2014,DarkwahOppong2019}, and the exploration of novel quantum phases such as exotic superfluidity \cite{Iskin2007}, interaction induced insulating phases \cite{Ospelkaus2006,Gunter2006,Taie2012b,Xu2013,Nataf2016,Lisandrini2017} and quantum magnetism \cite{Greif2013,Parsons2016,Mazurenko2017}.
	
	In particular, Fermi-Fermi mixtures have sparked experimental as well as theoretical interest, especially motivated by the analogy of fermionic pairing \cite{Demarco1999,Zwierlein2003,Greiner2003,Zwierlein2006,Schunck2007,Taglieber2008} and superfluid phenomena in condensed matter systems.

	Going beyond the typical SU(2) paradigm describing spin-1/2 electrons in the context of solid state systems, mixtures of high-spin ultracold fermions have been in the focus of research to investigate novel phenomena such as spin-changing dynamics \cite{Krauser2012,Krauser2014} or exotic SU($\mathcal{N}>2$) ground states, characterized by topological order and long-range quantum entanglement \cite{Honerkamp2004,Cazalilla2014a}.
	
	Fermi gases with SU($\mathcal{N}=2I+1$)-symmetric interactions as realized with alkaline-earth(-like) atoms such as Yb (nuclear spin $I=1/2$ and $I=5/2$) \cite{Fukuhara2007a,Taie2010} or Sr ($I=9/2$) \cite{Desalvo2010,Tey2010} allowed first groundbreaking experiments on thus far unexplored phenomena like SU($\mathcal{N}$) quantum magnetism in 1D \cite{Pagano2014}, the contact in a bulk SU(6) Fermi gas \cite{Song2019,Choudhury2020}  or effects of spin degeneracy like Pomeranchuk cooling in SU($ \mathcal{N} $) Mott insulators \cite{Honerkamp2004,Hermele2009,Xu2010,Taie2012a,Hofrichter2016a,Xu2018}. Isotope mixtures of SU($\mathcal{N}$)-symmetric Fermi gases \cite{Taie2010,Yip2011} characterized by a competition between inter- and intraisotope interactions hold promise to exhibit exotic $d$-wave superfluidity \cite{Lai2013} or two-flavor superfluid symmetry-locking phases \cite{Lepori2015,PintoBarros2017}.
	
	Alkaline-earth-like atoms further provide a long-lived metastable state, accessible via the doubly forbidden clock transition $^1$S$_0\rightarrow^3$P$_0$, which gives rise to interorbital spin- exchange interactions \cite{Scazza2014a,Cappellini2014b,Riegger2018,Ono2019a}. 
	These are expected  to allow studying systems with orbital degrees of freedom like the Kondo lattice model \cite{Gorshkov2010b,Foss-Feig2010,Foss-Feig2010a,Zhang2016a,Kanasz-Nagy2018b} or Kugel-Khomskii model \cite{Gorshkov2010b,Cazalilla2014a} and furthermore promise exotic interorbital superfluidity in the vicinity of an interorbital Feshbach resonance \cite{Zhang2015b,Hofer2015,Pagano2015,He2016}.
	
	In this article we report on the first measurements of interorbital interisotope interactions in Fermi-Fermi mixtures of $^{171}$Yb and $^{173}$Yb employing the ultranarrow clock transition.
	Our measurements strongly extend the knowledge about these Fermi-Fermi mixtures, thus enabling future studies of SU(2)$\otimes$SU($\mathcal{N}>2$)-symmetric systems.
	Especially the strong differences of the observed loss coefficients are crucial for further experiments.
	
	
	\section{Experimental Setup}
	\label{sec:methods}
	We start our experiments by trapping and cooling Yb atoms in a 2D-/3D-magneto-optical trap (MOT) operating on the $^1\text{S}_0$ $\rightarrow$ $ ^1\text{P}_1$ and the $^1\text{S}_0$ $\rightarrow$ $^3\text{P}_1$ transition, respectively \cite{Dorscher2013}.
	Subsequently, we use forced evaporative cooling in a two-color crossed optical dipole trap to create ultracold gases and mixtures of different Yb isotopes.
	The small intraisotope scattering length of $^{171}_{171}a_\text{gg}=-3(4)\,a_{0}$ prevents direct evaporative cooling of $^{171}$Yb$\,(I=1/2) $ to quantum degenaracy \cite{Kitagawa2008}.  
	We resolve this by adding $^{173}$Yb $\,(I=5/2) $ atoms.
	The corresponding intra- and interisotope $s$-wave scattering lengths $ ^{173}_{173}a_\text{gg} = 199 (2)~a_{0}$ and $^{171}_{173}a_\text{gg} = -580(60)\, a_{0}$ allow for efficient sympathetic cooling \cite{Kitagawa2008,Taie2010}.
	Here $ ^x_y a _{ij} $ denotes the scattering length between isotope $ x $ in the internal state $ i $ and isotope $ y $ in the internal state $ j $.
	The internal ground and excited states $^1$S$_0$ and $^3$P$_0$ are denoted by the subscripts `g' and `e', respectively. 
	We prepare arbitrary spin mixtures by applying a series of optical pump pulses on the $ ^1\text{S}_0 $ $ \rightarrow $ $ ^3 \text{P} _1 $ transition after a first evaporation stage.
	We  measure the population of individual $ m_\text{F}$ states using an optical Stern-Gerlach technique.
	For each experiment we adapt MOT loading times and evaporation ramps to optimize the particle number and temperature.
	Gases of $^{171}$Yb usually contain $N_{171} \approx (20\text{-}40) \cdot 10^3$ atoms at $ T_{171} \approx (0.2\text{-}0.3) T_{\text{F}}$, where $ T_{\text{F}} $ denotes the Fermi temperature.
	Mixtures of  $ ^{171} $Yb and  $ ^{173} $Yb consist of $ N_{171,173} \approx (10\text{-}40) \cdot 10^3$ at $ T_{171,173} \approx (0.25\text{-}0.55)\, T_{\text{F}}$.
	Subsequently, we adiabatically load the atoms into the lowest band of an optical lattice at the magic wavelength $\lambda_{\text{mag}}=759~$nm \cite{Lemke2009}.
	The lattice potential is composed of a triangular lattice in the vertical plane (referred to as the 2D lattice) and a 1D lattice in the perpendicular direction \cite{Becker2010}.
	We use momentum-resolved lattice modulation spectroscopy \cite{Heinze2011a} and interband clock spectroscopy to calibrate the 2D and 1D lattices.
	In the following the lattice depths are given in units of the recoil energy $ E_{\text{r}} = h \cdot 1.99 \, $kHz.
	We excite atoms from the $^1\text{S}_0$ ground state to the $^3\text{P}_0$ clock state using pulses of $\pi$-polarized clock laser light co-propagating with the 1D-lattice.
	Unless mentioned explicitly, we apply a magnetic field of $ B = 8.8\, \text{G} $ in the vertical direction to define a quantization axis.
	Throughout this paper we discuss excited state fractions $n_{\text{e}}= N_{\text{e}}/(N_{\text{g}}+N_{\text{e}}) $, (here $ N_{\text{e,g}} $ denotes the ground and excited state population) which we obtain by absorption imaging after time of flight.
	We employ repumping sequences on the $^3\text{P}_0$ $ \rightarrow $ $ ^3\text{D}_1$ transition and double-imaging techniques to detect the excited state atoms together with the remaining ground state population.

	\section{Interisotope Interactions}
	\label{sec:inter_isotope_interactions}
	
	To characterize the interisotope interaction, we prepare ground state isotope mixtures with $^{171}$Yb in the state $m_{F,171}$ and $^{173}$Yb in the state $m_{F,173}$ denoted by $|g,m_{F,171}; g,m_{F,173}\rangle$. 
	We use rectangular $\pi$-pluses with a duration $t_\text{pulse} = 1.6~\text{ms} \, (1.55~\text{ms} )$ for $^{171}$Yb ($^{173}$Yb), to excite either $^{171}$Yb ($^{173}$Yb) atoms to the $^3$P$_0$ state, resulting in a Fourier-limited spectroscopic line width of $625$~Hz.
	For each experimental sequence we determine the frequency of the blue sideband in the optical lattice to closely keep track of the actual 1D lattice depth.
	
	\begin{figure}
	  \includegraphics[width=0.45\textwidth]{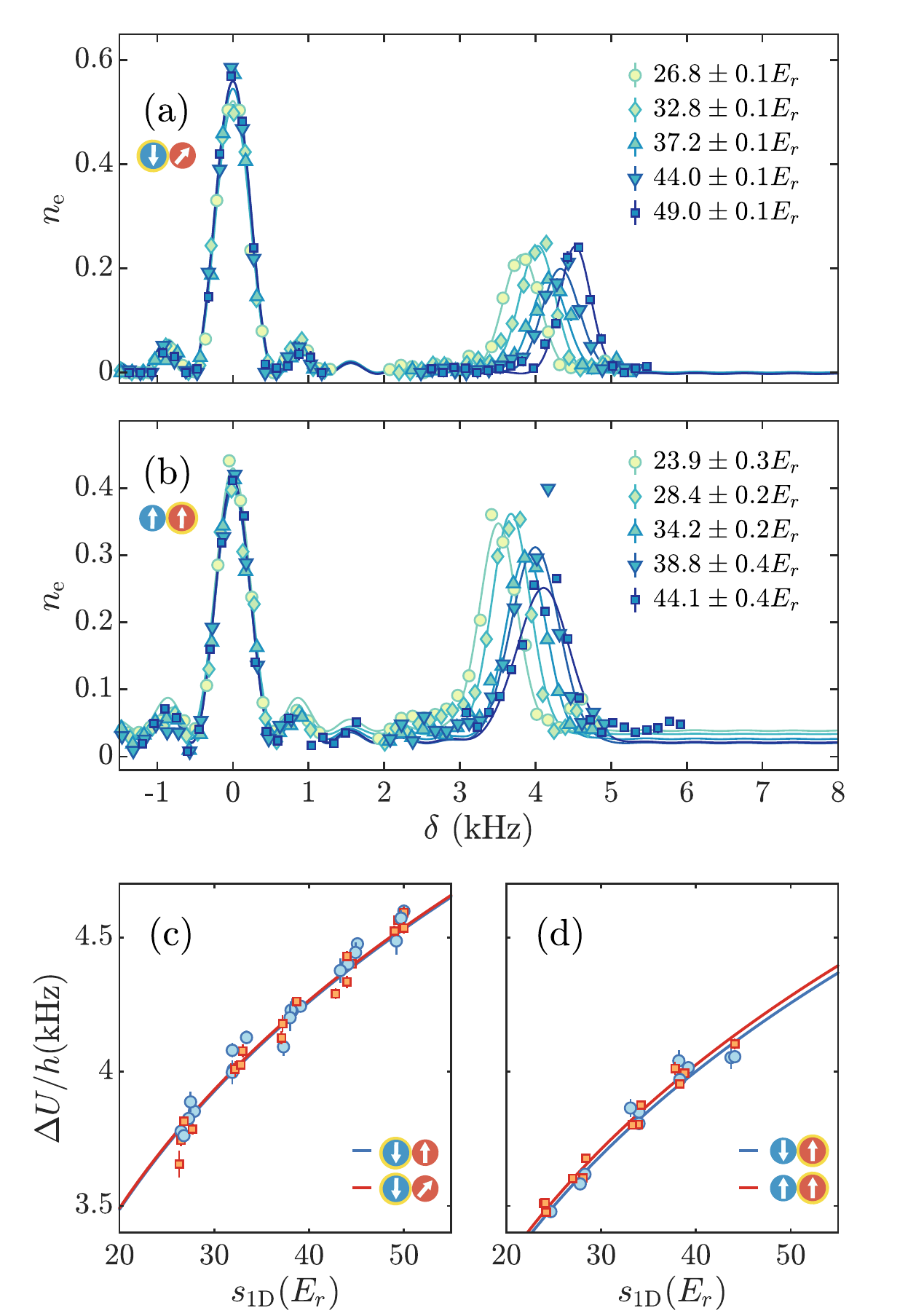}
	  \caption{
			Clock spectroscopy of doubly polarized $^{171}$Yb-$^{173}$Yb mixtures.
			(a) Spectroscopy on the $^{171}$Yb $^1$S$_0\rightarrow^3$P$_0$ transition for a mixture with $|\text{e},-1/2; \text{g},3/2 \rangle$.
			The excited state fraction $n_{\text{e}}$ is shown as a function of the clock laser detuning $ \delta $ with respect to the single atom transition.
			Colors (from bright to dark) indicate increasing 1D lattice depths.
			Every data point represents a single measurement. Solid curves represent best fits using a sinc$^2$ function for the singles feature and a gaussian for the interaction feature.
			(b) Interaction clock spectroscopy on the $^{173}$Yb $^1$S$_0\rightarrow^3$P$_0$ transition for a mixture with $ |\text{g},1/2; \text{e}, 5/2 \rangle$.
			(c) Interaction shift $ \Delta U $ for doubly occupied sites.
			Shown is the frequency difference between the $^{171}$Yb$_\mathrm{e}$-$^{173}$Yb$_\mathrm{g}$ doublon peak and $^{171}$Yb$_\mathrm{e}$ singles peak as obtained from the fits in (a) as a function of the 1D lattice depth. The 2D lattice is operating at $ s_{\text{2D}}=16.971(15)~E_\text{r} $. Different colors and symbols denote different spin configurations. Each data point represents a single spectrum with error bars representing fit uncertainties.   
			The solid lines are fits according to Eq. \ref{eq:U_of_Latdepth}. 
			(d) Same as (c) but for clock spectroscopy on the $^{173}$Yb $^1$S$_0\rightarrow^3$P$_0$ transition with $ s_{\text{2D}}=16.13(17)~E_\text{r}$.
	    }
	    \label{fig:inter_isotope_interaction_spectra}
	\end{figure}

	For a spin-polarized single-isotope Yb Fermi gas we expect a single spectroscopic feature at the bare atomic transition frequency, as the Pauli principle precludes double occupancy. 
	We use subsequent measurements of this unperturbed single-atom peak to incorporate a continuous a posteriori linear drift compensation of the clock laser frequency. 
	By adding atoms of a second isotope we allow for doubly occupied lattice sites and expect a second resonance peak, caused by interacting atom pairs and shifted by the corresponding differential on-site interaction energy: 
	\begin{equation}
	\Delta U =^{171}_{173}U_{\text{eg}}-^{171}_{173}U_{\text{gg}},
	\label{eq:delta_U}
	\end{equation}
	where $^x_y U_{ij}$ denotes the Hubbard on-site interaction given by: 
	\begin{equation}
	^{x}_{y}U_{ij}= \frac{4 \pi \, ^{x}_{y}a_{ij} \hbar^2 }{2\mu} \int \mathrm{d}\bm{r} \, | w_0(s_{\text{1D}},s_{\text{2D}},\bm{r})|^4. 
	\label{eq:U_of_Latdepth}
	\end{equation}
	Here $\mu$ denotes the reduced mass, $ \hbar $ the reduced Planck's constant, and $ w_0(s_{\text{1D}},s_{\text{2D}},\bm{r}) $ represents the single-particle Wannier function.
	Note that $w_0(s_{\text{1D}},s_{\text{2D}},\bm{r})$ depends on the lattice geometry and the lattice depths $s_\text{1D}$ and $s_\text{2D}$ of the 1D and 2D lattices.
	To underline the precision of our measurements we give values of $\Delta a= a_\text{eg}-a_\text{gg}$ throughout the paper to omit the large uncertainty in the literature value of $a_\text{gg} = -580(60)~a_0$ \cite{Kitagawa2008} in our reasoning of SU(2)$\otimes$SU(6) symmetry.
	
	Exemplary spectra for different values of the 1D lattice depths $ s_{\text{1D}} $ are shown in Fig.~\ref{fig:inter_isotope_interaction_spectra}(a) and (b).
	For each lattice depth we find a second spectroscopic feature shifted by a certain energy $ \Delta U $.
	While the single-particle peak is unaffected when changing the 1D lattice depth as expected, the second spectroscopic feature shifts according to an increasing Wannier integral, thus confirming the two-body interaction nature of the peak.
	Due to the close-to-perfect decoupling of nuclear spin and electronic angular momentum in the involved atomic states, all characterized by $ J=0 $, we expect the underlying interaction to be independent of the nuclear spin projection of the two interacting atoms.
	In the following, we argue that owing to the underlying symmetry of the contact interaction it is sufficient to compare only two specific pairs of atoms to unambiguously prove the aforementioned SU(2)$ \otimes $SU(6) symmetry.
	
	The Hamiltonian of the underlying rotationally symmetric $s$-wave interaction commutes with the total spin operator $ F^2 $ of the two colliding atoms.
	Here $ \vec{F}=\vec{f}_1 + \vec{f}_2 $ and the total magnetization is $ M = m_1 + m_2 $. The interaction Hamiltonian $H_\text{int}$ can thus be conveniently written in terms of eigenfunctions $ |F,M \rangle $ of $ F^2 $:
	\begin{equation}
	H_{\text{int}}=\frac{4\pi \hbar}{2\mu}\delta(\vec{r}_1-\vec{r}_2)\sum_{F=F_{\text{min}}}^{F_{\text{max}}} \sum_{M=-F}^{F}a_{F}|F,M\rangle \langle F,M|.
	\end{equation}
	Here $ \delta(\vec{r}_1-\vec{r}_2) $ denotes the Dirac $ \delta $-function for atoms at positions $ \vec{r}_1 $ and $ \vec{r}_2 $, $ a_{F}$ describe the $F$-dependent scattering lengths, which in general differ for scattering channels with different total spin $ F $.
	In our experiments however we observe atoms in the magnetization basis $|m_1,m_2\rangle$ that can be obtained from the interaction basis through the basis transformation
	\begin{equation}
	|m_1,m_2\rangle =\sum_{m_1+m_2=M} c_{m_1,m_2}^{F,M}  |F,M\rangle,
	\label{eq:magnetization basis}  
	\end{equation}  
	where $m_1$ ($ m_2 $) denotes the spin state of $^{171}$Yb ($^{173}$Yb).
	The summation is carried out over all combinations $ (m_1,m_2) $ with $ m_1+m_2=M $ and $ c_{m_1,m_2}^{F,M} $ denote the Clebsch-Gordan coefficients. 
	In the specific case of $^{171}$Yb-$^{173}$Yb Fermi-Fermi mixtures, where $ f_1 = 1/2 $ and $ f_2 = 5/2$, the allowed scattering channels carry total spin $ F=2 $ and $ F=3 $. 
	According to Eq. \ref{eq:magnetization basis} it is thus sufficient to measure the interaction shift of any two spin configurations to obtain $a_2$ and $a_3$, thereby testing the $\text{SU}(2)\otimes\text{SU}(6))$  symmetry.
	
	First, we investigate the interaction parameters $^{171}_{173}U_\text{eg}$ (compare Fig.~\ref{fig:inter_isotope_interaction_spectra}(a) and \ref{fig:inter_isotope_interaction_spectra}(c)).
	To access these, we excite $^{171}$Yb atoms and study the spectra of the two different spin configurations (compare Eq. \ref{eq:magnetization basis})
	\begin{align}
	|-1/2 , 5/2\rangle &= \sqrt{\frac{1}{6}}|3,2\rangle +\sqrt{\frac{5}{6}} |2,2\rangle  \label{eq:171 Spin config 2}  		\\
	|-1/2 , 3/2\rangle &= \sqrt{\frac{1}{3}} |3,1\rangle + \sqrt{\frac{2}{3}} |2,1\rangle
	\label{eq:171 Spin config 2}  		
	\end{align}
	to test the symmetry properties of the interaction.
	We find $ \Delta a_{-1/2,5/2}= 495(1)~a_0 $ and $ \Delta a_{-1/2,3/2}= 496(2)~a_0 $ corresponding to $\Delta a_2 =494(4)~a_0$ and $\Delta a_3=499(9)~a_0$  consistent with the assumption of $\text{SU}(2)\otimes\text{SU}(6)$  symmetry,
	with an average scattering length $\Delta \bar{a}_\text{eg} = 495(2)~a_0$.
	
	\begin{center}
		\begin{table}
			\caption{Summary of all interaction parameters measured in this work.
			The subscripts $i$ and $j$ in $a_{i,j}$ and $\beta_{i,j}$ represent the $m_F$ state of the $^{171}$Yb and $^{173}$Yb isotopes, respectively.
			The quantities $\bar{a}$ and $\bar{\beta}$ are the best estimate for the elastic scattering length and the loss coefficient, respectively.
				\label{table:summary}}
			\begin{tabular}{|c | c | c |} 
				\hline
				Quantity & $^{171}$Yb$_\text{e}$-$^{173}$Yb$_\text{g}$ & $^{171}$Yb$_\text{g}$-$^{173}$Yb$_\text{e}$ \\ [0.5ex] 
				\hline\hline
				$\Delta a_{-1/2,+5/2}$ ($a_0$) & $495 \pm 1 $ 				& $479.1 \pm 1.5$ \\ 
				\hline
				$\Delta a_{-1/2,+3/2}$ ($a_0$) & $496 \pm 2 $ 				& -  \\
				\hline
				$\Delta a_{+1/2,+5/2}$ ($a_0$) & - 					& $482 \pm 1 $  \\
				\hline
				$\Delta a_2$ ($a_0$) 		& $494 \pm 4 $ 				& $478.5 \pm 1.8$  \\
				\hline
				$\Delta a_3$ ($a_0$) 		& $499 \pm 9 $ 				& $482.1 \pm 1.1$  \\
				\hline
				$\Delta \bar{a}$ ($a_0$)	& $495 \pm 2 $ 				& $481.2 \pm 0.9$  \\
				\hline
				$\beta_{-1/2,+5/2}$ (cm$^3$s$^{-1}$)	& $(1.69 \pm 0.07)\cdot 10^{-12}$  	& $(4.6 \pm 1.7)\cdot 10^{-15}$ \\ 
				\hline
				$\beta_{-1/2,+3/2}$ (cm$^3$s$^{-1}$)	& $(1.79 \pm 0.05)\cdot 10^{-12}$  	& -  \\
				\hline
				$\beta_{+1/2,+5/2}$ (cm$^3$s$^{-1}$)	& - 					& $(3 \pm 1)\cdot 10^{-15}$  \\
				\hline
				$\bar{\beta}$ (cm$^3$s$^{-1}$)	& $(1.75 \pm 0.04)\cdot 10^{-12}$	& $(3.6 \pm 0.9)\cdot 10^{-15}$  \\
				\hline
			\end{tabular}
		\end{table}
	\end{center}	
	
	Next, we study the interaction parameters $^{171}_{173}U_\text{ge}$. For this purpose we excite $ ^{173}$Yb in the two different spin configurations
	\begin{align}
	|1/2 , 5/2\rangle &= |3,3\rangle \\
	|-1/2 , 5/2\rangle &= \sqrt{\frac{1}{6}} |3,2\rangle + \sqrt{\frac{5}{6}} |2,2\rangle
	\label{eq:173 Spin config}  	
	\end{align}
	and obtain $\Delta a_{1/2,5/2}= 482(1)\, a_0$ and $ \Delta a_{-1/2,5/2}=497.1(1.5)\,a_0 $ and correspondingly $ \Delta a_2=478.5(1.8)\,a_0 $  and $ \Delta  a_3=482.1(1.1)\,a_0 $.
	Again, within our experimental uncertainties these are equal, confirming $ \text{SU}(2)\otimes\text{SU}(6)$ symmetry, with an average scattering length $\Delta \bar{a}_\text{ge} = 481.2(9)~a_0$.
	
	Note that the scattering lengths for the two different interorbital interisotope mixtures $^{171}$Yb$_\mathrm{e}$-$^{173}$Yb$_\mathrm{g}$ and $^{171}$Yb$_\mathrm{g}$-$^{173}$Yb$_\mathrm{e}$ are very similar, yet not identical. 
	While different scattering lengths of different isotope mixtures, e.g. Yb \cite{Kitagawa2008},
	can often be well-explained using a mass-scaling model, this does not apply for the interorbital mixtures we investigate, because they have the same reduced mass. 
	Therefore, our measurements could prove to be an interesting benchmark for beyond-mass scaling contributions,
	which are required to explain the difference between the $^{171}$Yb$_\text{e}$-$^{173}$Yb$_\text{g}$ and $^{171}$Yb$_\text{g}$-$^{173}$Yb$_\text{e}$ scattering lengths at the same reduced mass.
	
	For the absolute values of the scattering lengths we find $^{171}_{173}\bar{a}_{eg}=-85(60)a_0$ and $^{171}_{173}\bar{a}_{ge}=-99(60)a_0$, using the literature value of $^{171}_{173}$Yb$_\text{gg}$ \cite{Kitagawa2008},

\section{Interaction-Induced Losses}
\label{sec:losses}
To characterize the inelastic interactions of both inter-orbital interisotope mixtures, we excite atom pairs at the corresponding resonance frequencies as obtained in the previous section and measure the excited state population as a function of lattice hold time and determine the decay rates.
In this way, we measure the number of interacting pairs, which is expected to exhibit an exponential decay \cite{Goban2018}.
In analogy to the previous section, we check for the anticipated SU(2)$\otimes$SU(6) symmetry of the inelastic interactions by performing all measurements described below for two different spin configurations.

\begin{figure}
	\includegraphics[width=0.45\textwidth]{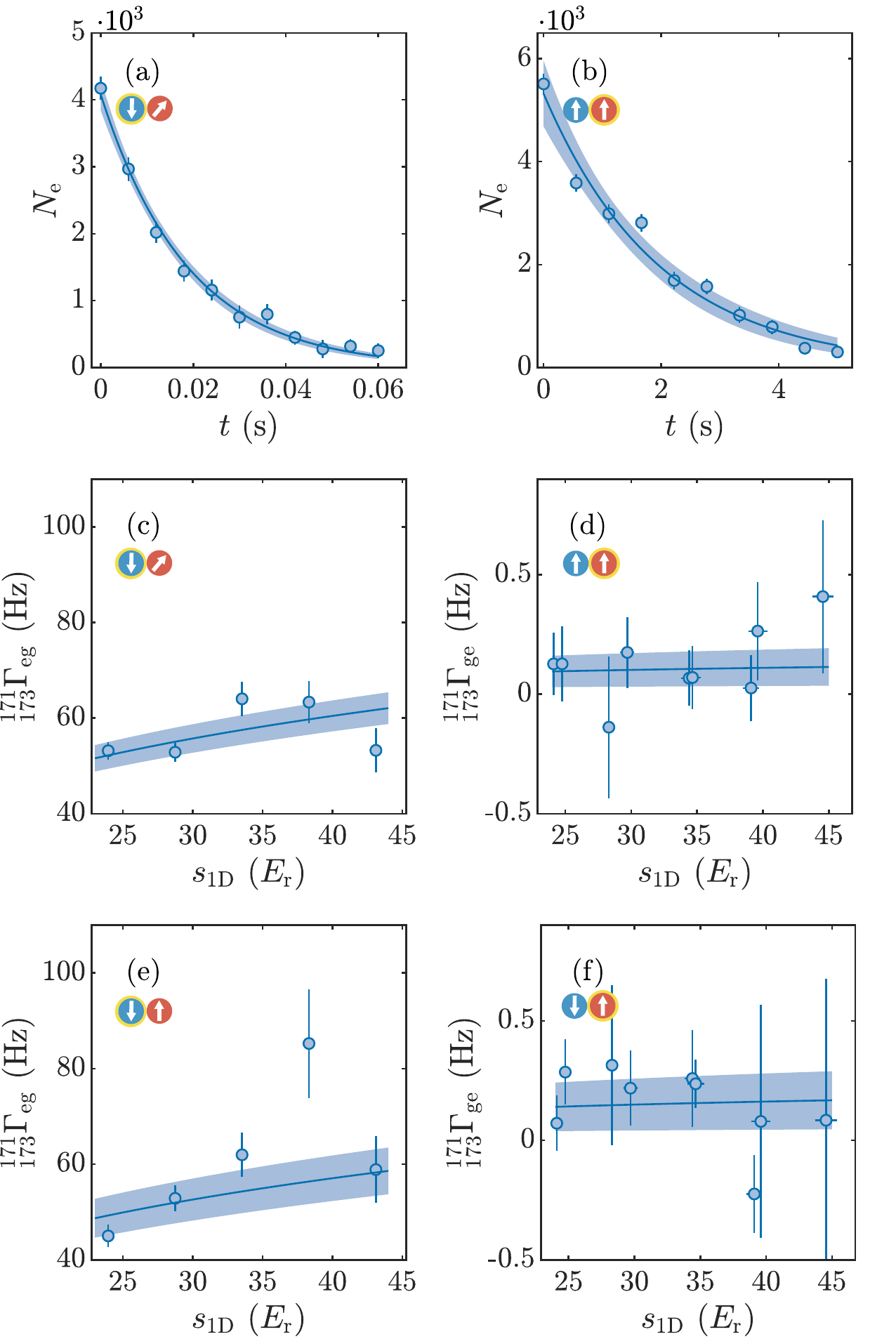} 
	\caption{
		Losses of doubly polarized $^{171}$Yb-$^{173}$Yb mixtures.
		(a) Atom number of $^{171}$Yb$_{\text{e}}$ atoms of interacting $|e, -1/2; g, 5/2\rangle$ pairs as a function of hold time $t$.
		Here $s_{1\text{D}} = 24.0(2)~E_\text{r}$ and $s_{2\text{D}} = 16.0(1)~E_\text{r}$.
		Error bars depict the experimental atom number uncertainty.
		The solid curve is an exponential fit and the shaded area is the 95\% confidence interval.
		(b) Same as in (a), but for $|g, 1/2; e, 5/2\rangle$ pairs.
		Here $s_{1\text{D}} = 34.7(6)~E_\text{r}$ and $s_{2\text{D}} = 17.0(4)~E_\text{r}$.
		(c) and (e) show the decay rates $^{171}_{173}\Gamma_{\text{e,g}}$ of the different spin configurations of $|e, -1/2; g, 5/2\rangle$ and $|e, -1/2; g, 3/2\rangle$ pairs, respectively.		
		Each data point is averaged several times and error bars indicate the propagated fit uncertainties.
		The solid line shows a fit according to Eq.~\ref{eq:losses} to determine $^{171}_{173}\beta_{\text{e,g}}$ (see Table~\ref{table:summary}).
		(d) and (f) are the same as in (c) and (e) but for $|g, 1/2; e, 5/2\rangle$ and $|g, -1/2; e, 5/2\rangle$ pairs, respectively.
	}
	\label{fig:losses}
\end{figure}

Figure \ref{fig:losses}(a) shows a typical measurement where $|e, -1/2; g, 5/2\rangle$.
We fit an exponential $N_0 \exp(-t \Gamma_\text{int})$, where the initial particle number $N_0$ and decay rate $\Gamma_\text{int}$ are free parameters, and obtain a decay rate for every $s_{1\text{D}}$.
We additionally measure the non-interacting single atom decay rate of $^{171}$Yb$_\text{e}$ by exciting on the bare atomic resonance, and similarly find the decay rate due to one-body losses $\Gamma_\text{one}$ \cite{footnote1}.
The decay rate due to inelastic collisions is then given by $^{171}_{173}\Gamma_{\text{e,g/g,e}} = \Gamma_{\text{int}} - \Gamma_{\text{one}}$.


$^{171}_{173}\Gamma_{\text{e,g/g,e}}$ is lattice depth dependent and is connected to the two-body loss coefficient $^{171}_{173}\beta_{\text{e,g/g,e}}$ by \cite{Goban2018}:
\begin{equation}
^{171}_{173}\Gamma_{\text{e,g/g,e}} = ^{171}_{173}\beta_{\text{e,g/g,e}} \int \text{d}\bm{r} |w_0(s_{\text{1D}},s_{\text{2D}},\bm{r})|^4 .
\label{eq:losses}
\end{equation}
We conduct the measurement shown in Fig.~\ref{fig:losses}(a) for different $s_{1\text{D}}$ and two different spin state configurations. 
The results for $|e, -1/2; g, 5/2\rangle$ ($|e, -1/2; g, 3/2\rangle$) are shown in Fig.~\ref{fig:losses}(c)((e)) and we find $^{171}_{173}\beta_{\text{e,g}} = (1.79 \pm 0.05) \cdot 10^{-12}~\mathrm{cm^3 s^{-1}}$ ($^{171}_{173}\beta_{\text{e,g}} = (1.69 \pm 0.07) \cdot 10^{-12}~\mathrm{cm^3 s^{-1}}$).
Therefore, we conclude that, within our measurement accuracy, we observe SU(2)$\otimes$SU(6) symmetry for the inelastic part of the $^{171}$Yb$_\text{e}$-$^{173}$Yb$_\text{g}$ interaction, characterized by an average loss coefficient $\bar{\beta}_\text{eg} = 1.75(4)\cdot 10^{-12}$~cm$^3$s$^{-1}$. 

We repeat this measurement for the $^{171}$Yb$_\text{g}$-$^{173}$Yb$_\text{e}$ interaction.
Figure~\ref{fig:losses}(b) shows a typical loss measurement on the interaction peak.
We determine the decay rates $^{171}_{173}\beta_{\text{g,e}}$ for the $|g, 1/2; e, 5/2\rangle$ ($|g, -1/2; e, 5/2\rangle$) mixture, shown in Fig.~\ref{fig:losses}(d)((f)) and obtain $^{171}_{173}\beta_{\text{e,g}} = (3.1 \pm 1.1) \cdot 10^{-15}~\mathrm{cm^3 s^{-1}}$ ($^{171}_{173}\beta_{\text{e,g}} = (4.6 \pm 1.7) \cdot 10^{-15}~\mathrm{cm^3 s^{-1}}$).
This confirms the SU(2)$\otimes$SU(6) symmetry for the inelastic part of the $^{171}$Yb$_\text{g}$-$^{173}$Yb$_\text{e}$ interaction, characterized by an average loss coefficient $\bar{\beta}_\text{eg} = 3.6(9)\cdot 10^{-15}$~cm$^3$s$^{-1}$. 
Note that the $^{171}$Yb$_\text{g}$-$^{173}$Yb$_\text{e}$ interaction decay rate is comparable to the one-body decay rate in the optical lattice, thus contributing significantly to the uncertainty of $^{171}_{173}\Gamma_{\text{g,e}}$. 

Remarkably, as one of the main results of this paper we find that the inelastic interactions of the two different interorbital interisotope mixtures differ by more than two orders of magnitude, even though the elastic interactions are very similar, making $^{171}$Yb$_\text{g}$-$^{173}$Yb$_\text{e}$ the more promising candidate for quantum simulations.

\section{Spin-Exchange Spectroscopy of $^{171}$Yb}
\label{sec:spin_exchange_171}

\begin{figure}
	\includegraphics[width=0.45\textwidth]{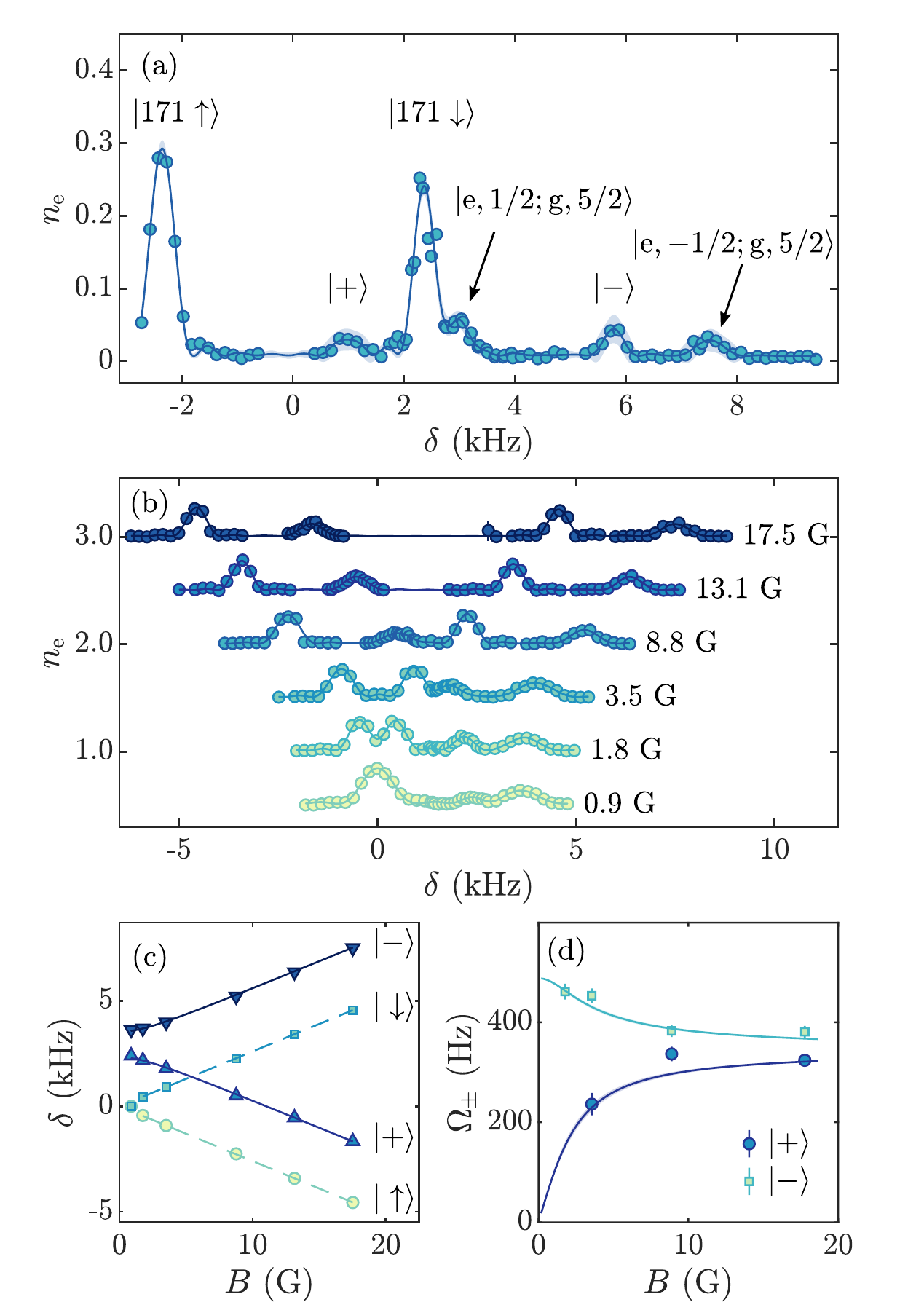} 
	\caption{
		Spin-exchange interactions in $^{171}$Yb.
		(a) High resolution Fourier-limited clock spectroscopy of a mixture of spin balanced $^{171}$Yb and spin-polarized $^{173}$Yb at $B = 8.8$~G.
		Every data point is averaged three times with error bars indicating one standard deviation.
		The solid curve is a fit as obtained using six sinc$^2$ functions to guide the eye.
		The shaded area represents the 95\% confidence interval of the fit.
		(b) High-resolution Fourier-limited clock spectroscopy of a spin balanced $ ^{171} $Yb gas.
		We plot the excited state fraction $n_{\text{e}}$ as a function of the detuning relative to the center frequency of the two single peaks.
		Data sets for different magnetic fields are offsetted in the $y$-direction for clarity in ascending order.
		Every data point is averaged three times with error bars indicating one standard deviation.
		Solid curves are fits as obtained using several sinc$^2$ functions with a small offset.
		(c) Peak position relative to the zero field singles clock frequency as a function of magnetic field, as obtained from fits in (a).
		The error bar is the fit uncertainty (usually smaller than the point size).
		Circles (squares) represent the spin up (down) single peaks.
		The dashed lines are fits to the corresponding Zeeman shift.
		Triangles up (down) represent the $ |+\rangle $ ($ |-\rangle $) peaks.
		The solid lines are fits according to Eq.~\ref{eq:f_dbl} to determine the $  U_{\text{eg}^+} $ and $  U_{\text{eg}^-} $ interaction.
		We obtain $\Delta U_{\text{eg}^+}/h=3.53(4)$~kHz and $\Delta U_{\text{eg}^-}/h=2.32(3)$~kHz. 
		(d) Rabi frequencies for excitation of the  $ |+\rangle $ (circles) and  $ |-\rangle $ (squares) states as a function of magnetic field.
		We obtain the Rabi frequencies by fitting Rabi oscillations on both interaction peaks.
		The solid line represents the calculated $\Omega_{\pm}(B)$ for the value of $V_\text{ex}$ that we found in (c).
		The shaded area shows the 95\% confidence interval (barely visible).}
	\label{fig:full_mixture_and_171_spin_exchange}
\end{figure}

In the following we extend complexity and prepare a spin-balanced gas of $ ^{171} $Yb together with a spin-polarized gas of $ ^{173} $Yb in a deep optical lattice with $s_\text{1D} = 40~E_\text{r}$ and $s_\text{1D} = 25~E_\text{r}$. 
We excite $^{171}$Yb and observe six spectroscopic features as shown in Fig.~\ref{fig:full_mixture_and_171_spin_exchange}(a).
We identify the two large peaks as being caused by sites singly occupied with either $^{171}{\text{Yb}}_{\uparrow}$ or $^{171}{\text{Yb}}_{\downarrow}$ particles.
They are separated by the Zeeman energy $\Delta E_{\text{Z}} = -399.0(1)~\text{Hz/G} \cdot \Delta m_F B$ \cite{Bettermann2020a}.
In addition, we observe two interisotope interorbital peaks shifted by $\Delta ^{171}_{173}U_{\text{eg}}$ relative to the $^{171}{\text{Yb}}_{\uparrow}$ and $^{171}{\text{Yb}}_{\downarrow}$ states, respectively.
The two remaining peaks, indicated by $|\pm\rangle$, can be attributed to doubly occupied lattice sites of $^{171} $Yb with one excitation present in the pair.
These peaks can be understood as follows (see also Refs.~\cite{Ono2019a,Bettermann2020a}).
At vanishing magnetic field the anti-symmetric wave function  required for two indistinguishable fermionic particles can be realized in two ways: either by a spin singlet with a symmetric orbital wave function denoted by
$|\text{eg}^{+} \rangle =  1/\sqrt{2}  (|\text{eg}\rangle +|\text{ge}\rangle) \otimes |\text{s}\rangle$
or a spin triplet with an anti-symmetric orbital wave function denoted by 
$ |\text{eg}^{-} \rangle = 1/\sqrt{2}  (|\text{eg}\rangle -|\text{ge}\rangle) \otimes  |\text{t}\rangle $.
Each of these two interaction channels is characterized by its individual molecular potential and thus individual on-site interaction $ U_{\text{eg}^\pm} $ and scattering length $ a _{\text{eg}^\pm}$.
At finite magnetic field the $ |\text{eg}^{+} \rangle $ and $ |\text{eg}^{-} \rangle $ states mix.
The new eigenstates can be written as 
\begin{eqnarray}
	|\pm\rangle = c_1(B)|eg^\pm\rangle\pm c_2(B)|eg^\mp\rangle 
\end{eqnarray}
with corresponding eigenenergies 
\begin{equation}
E_{\pm} = V \mp \sqrt{V_{\text{ex}}^2+E_{\text{Z}}^2}.
\label{eq:f_dbl}
\end{equation}
Here $ V=(U_{\text{eg}^+}+U_{\text{eg}^-})/2 $ denotes the direct part and  $ V_{\text{ex}}=(U_{\text{eg}^+}-U_{\text{eg}^-})/2 $ the spin-exchange part of the interaction, while $E_\text{Z}$ is the Zeeman energy.
The mixing coefficients $ c_{1,2}(B) $ are given by:
 \begin{eqnarray}
 c_1(B)=\frac{V_{\text{ex}}+\sqrt{V_{\text{ex}}^2+E_{\text{Z}}^2(B)}}
 {\sqrt{2V_{\text{ex}}^2+2E_{\text{Z}}^2(B)+2V_{\text{ex}}\sqrt{ V_{\text{ex}}^2 +E_{\text{Z}}^2(B)}}}\\
  c_2(B)=\frac{E_{\text{Z}}(B)}
 {\sqrt{2V_{\text{ex}}^2+2E_{\text{Z}}^2(B)+2V_{\text{ex}}\sqrt{ V_{\text{ex}}^2 +E_{\text{Z}}^2(B)}}}.
 \end{eqnarray}
At vanishing magnetic field, $ c_2$ equals zero and therefore the $ |+\rangle $ state decouples and cannot be excited.

In general, the coupling is determined by the Rabi frequencies of the two particle transitions, given by:
 \begin{eqnarray}
 \Omega_{\pm}(B) =\sqrt{2} \, c_{2,1}(B) \Omega_{\text{s}},
 \label{eq:IntRabiFreq}
 \end{eqnarray} 
where $ \Omega_{\text{s}} $ denotes the Rabi frequency of singly occupied sites.
To achieve an optimal spectroscopic contrast during the measurement we account for the super- and subradiant properties of the $|\text{eg}^{-} \rangle $ and $|\text{eg}^{+} \rangle $ states by adapting the length of the clock pulse to fulfil the $ \pi $-pulse condition for each individual spectroscopic feature.

To determine the direct interaction $ V $ and the exchange interaction $V_{\text{ex}}$ we prepare a spin-balanced Fermi gas of $ ^{171}$Yb with $N\approx(15-35)\cdot 10^3$ atoms and perform spectroscopy in a deep lattice ($s_\text{1D}=50~E_\text{r}$ and $s_\text{2D}=25~E_\text{r}$) at different magnetic fields.
The resulting spectra are shown in Fig.~\ref{fig:full_mixture_and_171_spin_exchange}(b). 
We fit these spectra with a multi-peak function and obtain the resonance frequencies, that are depicted in Fig.~\ref{fig:full_mixture_and_171_spin_exchange}(c).
We find four spectroscopic branches, two belonging to the single-particle transitions denoted by $ |\uparrow\rangle$ and $ |\downarrow \rangle $, which are shifted by the Zeeman energy $ E_{\uparrow\downarrow} =\pm E_{\text{Z}}(B)$.
We use these single-body transitions to calibrate the magnetic field.
The remaining two branches correspond to an excitation of an interacting atom pair $|\pm\rangle$ at energy $E_\pm$.
We fit Eq.~\ref{eq:f_dbl} to the resonance positions to obtain the energy shifts $ \Delta U_{\text{eg}^\pm}=U_{\text{eg}^\pm}-U_{\text{gg}}$.
From our measurement we find $ \Delta U_{\text{eg}^+}/h=3.53(4)$~kHz and $ \Delta U_{\text{eg}^-}/h=2.32(3)$~kHz corresponding to $s$-wave scattering lengths of $ a_{\text{eg}^+}=203(5)~a_0$ and $ a_{\text{eg}^-}=308(6)~a_0$, respectively, in fair agreement with previously measured values \cite{Ono2019a, Bettermann2020a}.
We explicitly measure the Rabi frequencies of the two particle transitions at different magnetic fields, shown in Fig.~\ref{fig:full_mixture_and_171_spin_exchange}(d)
Our results are in good agreement with the spectroscopically measured value of $V_\text{ex}$ (compare Eq.~\ref{eq:IntRabiFreq}).

The interorbital spin-exchange interaction $ V_{\text{ex}} $ has a negative sign and thus is antiferromagnetic, which makes $ ^{171} $Yb a promising candidate for quantum simulation of the `Kondo lattice model' \cite{Gorshkov2010b,Foss-Feig2010,Foss-Feig2010a,Zhang2016a,Kanasz-Nagy2018b}.

\section{Conclusion \& Discussion}
\label{sec:conclusion_discussion}

In conclusion, we have characterized the elastic and inelastic part of interisotope interorbital interactions in different mixtures of $^{171}$Yb-$^{173}$Yb, summarized in Table~\ref{table:summary}.
Our measurements directly show SU(2)$\otimes$SU(6) symmetry for both elastic and inelastic interactions in $^{171}$Yb$_\text{e}$-$^{173}$Yb$_\text{g}$ and $^{171}$Yb$_\text{g}$-$^{173}$Yb$_\text{e}$ mixtures. 
The elastic interactions are similar, which we attribute to the high degree of symmetry of the underlying molecular potentials. 
Together with the data obtained in \cite{Scazza2014a,Cappellini2014b,Hofer2015,Pagano2015,Bouganne2017,Franchi2017,Cappellini2019,Bettermann2020a} our measurements could contribute to a more precise understanding of the molecular potential of excited-ground state dimers in Yb.
Additionally, the inelastic interactions differ by more than two orders of magnitude, which could prove interesting as a test for future theoretical models. 


The interorbital mixtures we investigated could be better-suited than the proposed ground state mixture to reach two-flavor superfluid symmetry-locking phases \cite{Lepori2015,PintoBarros2017} since it is deeper in corresponding regime, likely allowing for less demanding temperatures. 

Furthermore, we measured the interorbital spin-exchange interactions of $^{171}$Yb, finding $a_\text{eg+} = 203(5) a_0$ and $a_\text{eg-} = 308(6) a_0$ in fair agreement with Refs.~\cite{Ono2019a,Bettermann2020a}.
The negative $V_\text{ex}$ shows that ${^{171}}$Yb is a promising platform to realize the antiferromagnetic Kondo-lattice model \cite{Gorshkov2010b,Foss-Feig2010,Foss-Feig2010a,Zhang2016a,Kanasz-Nagy2018b}.



\acknowledgements
This work has been funded by the Deutsche Forschungsgemeinschaft (DFG, German Research
Foundation) - SFB-925 - project 170620586.
We thank L. Mathey, L. Freystatzky, P. Schmelcher, S. Mistakidis and G. Bougas for useful discussions.

B.~Abeln and K.~Sponselee contributed equally to this work.

\newpage

\appendix{}
\clearpage

\providecommand{\noopsort}[1]{}\providecommand{\singleletter}[1]{#1}%


\begin{thebibliography}{83}
\bibitem [{foo()}]{footnote1}%
  \BibitemOpen
  \href@noop {} {\bibinfo  {journal} {{We do this because in this measurement
  interacting excited Yb atoms can be lost by two processes, either one-body
  losses in exactly the same way as non-interacting excited atoms, or by
  inter-isotope inelastic collisions}}\ }\BibitemShut {NoStop}%
\end{thebibliography}
\end{document}